\documentclass{elsart}

\usepackage{natbib}

\usepackage{amssymb}

\usepackage{graphicx}

\begin{document}

\begin{frontmatter}



\author[lanl]{R.~Reifarth\corauthref{cor}},
\ead[url]{www.reifarth.com}
\ead{reifarth@lanl.gov}
\author[lanl]{R.C.~Haight},
\author[fzk]{M.~Heil},
\author[fzk]{F.~K\"appeler},
\author[lanl]{D.J.~Vieira}
\address[lanl]{Los Alamos National Laboratory, Los Alamos, New Mexico, 87545, USA}
\address[fzk]{Forschungszentrum Karlsruhe, Institut f\"{u}r Kernphysik,
       Postfach 3640, D-76021 Karlsruhe, Germany}

\corauth[cor]{Corresponding author:}

\title{Neutron capture measurements at a RIA-type facility}

\begin{abstract}

Neutron capture cross sections of unstable isotopes are important for 
neutron induced nucleosynthesis as well as for technological applications.
The Rare Isotope Accelerator (RIA) or comparable facilities will be able 
to produce radioactive ion beams up to 10$^{12}$ 
particles/s and would therefore be a suitable place for (n,$\gamma$) studies on 
radioactive isotopes with half-lives between days and months. 
We propose a facility for measurements of (n,$\gamma$) cross sections of unstable 
isotopes in the keV range suited for minimal sample masses down to 10$^{15}$ atoms, 
corresponding to minimum half-lives of only 10~d.

\end{abstract}


\end{frontmatter}

\section{Introduction}
\label{}

In astrophysics the neutron energy range between 1~keV and 1~MeV is most important, 
because 
it corresponds to the temperature regimes of the relevant sites for synthesizing 
all nuclei between iron and the actinides. In this context (n,$\gamma$) cross sections 
for unstable isotopes are requested for the s-process related to stellar helium 
burning as well as for the r- and p-processes related to explosive nucleosynthesis 
in supernovae. In the s process, these data are required for analyzing 
branchings in the reaction path, which can be interpreted as diagnostic tools for 
the physical state of the stellar plasma. Most of the nucleosynthesis reactions during 
the r and p processes occur outside the stability valley, involving rather 
short-lived nuclei. Here, the challenge for (n,$\gamma$) data is linked to the freeze-out 
of the final abundance pattern, when the remaining free neutrons are captured as 
the temperature drops below the binding energy. Since 
many of these nuclei are too short-lived to be accessed by direct measurements it 
is, therefore, essential to obtain as much experimental information as possible 
off the stability line in order to assist theoretical extrapolations of nuclear 
properties towards the drip lines. 

Apart from the astrophysical motivation there is continuing interest on neutron 
cross sections for technological applications, i.e. with respect to the neutron 
balance in advanced reactors, which are aiming at high burn-up rates, as well as 
for concepts dealing with transmutation of radioactive wastes.

The Rare Isotope Accelerator (RIA) \cite{Sav01} will produce radioactive 
ion beams up to 10$^{12}$ particles/s and would, therefore, be the ideal place for neutron 
capture studies on short-lived radioactive isotopes with half-lives between days and 
months. Conducting  this kind of experiments directly at RIA eases many technical 
and organizational problems, in particular the depletion of enriched samples by decay 
in transport, but also safety issues as well as sample handling. 

\section{Simulations}

All the discussed processes were aided by detailed Monte-Carlo simulations. The program 
used was the GEANT 3.21 package \cite{GEA93}, developed at CERN in combination with the GCALOR interface
\cite{ZeG94}, which is especially suited for tracking neutrons of thermal 
energies up to the MeV range.
The simulations including the BaF${_2}$ crystal ball are based on the DANCE array,
a soccer ball design with 162 crystals to cover the entire solid angle \cite{HSD79,WCD02}.
As it will become clear in later sections, one crystal needs to be left out to 
leave space for a proton pipe, which means that all the simulations contained have been 
carried out with 161 crystals. 
More details on the simulation technique can be found in Refs. \cite{HRK99,HRF01,RHK01}. 

\section{Experimental Approach}

Measurements on radioactive isotopes represent a stringent challenge for further 
improvements of experimental techniques. This holds for the neutron sources as 
well as for the detection systems. Though the activation method or accelerator 
mass spectroscopy of the reaction products could be applied in a limited number 
of cases (see e.g. \cite{RAH03, BBB02}), this report concentrates on the more universal method of detecting the 
prompt capture gamma-rays, which is required for the application of  neutron 
time-of-flight (TOF) techniques. 

In this paper we discuss the possibility for using an optimized low energy accelerator 
as a neutron source for a "Neutron capture program" at RIA-type facilities in combination 
with a highly segmented, high-efficiency $\gamma$-ray calorimeter.

\subsection{Neutron Source}

Since spallation or photo-neutron sources require large accelerators, it is assumed 
that a small accelerator is best suited for neutron experiments at RIA. This solution 
has the additional advantage that the neutron spectrum can be tailored to the specific 
energy range of interest.

\begin{figure}
\begin{center}
\includegraphics[width=12cm]{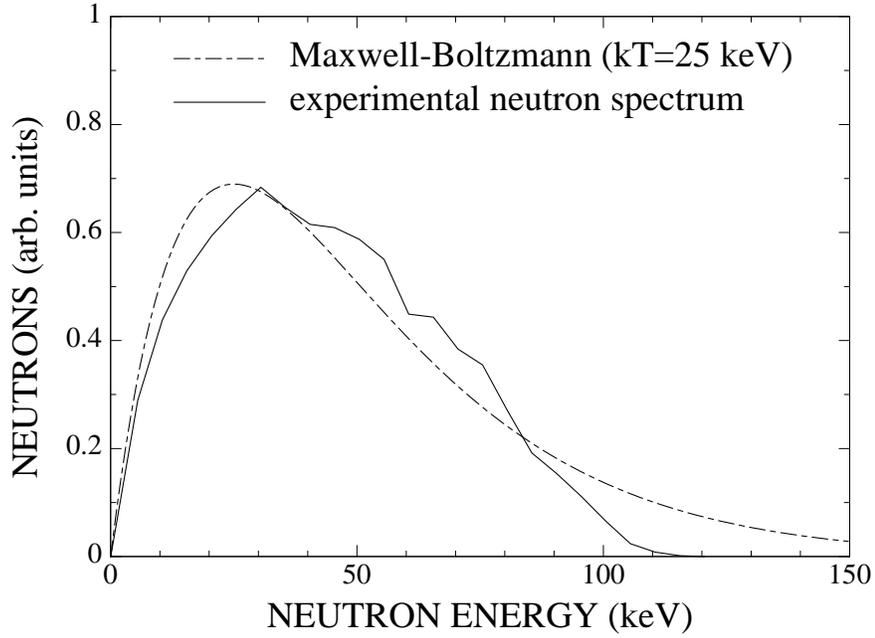}
\end{center}
\caption{The neutron spectrum resulting from bombarding of a metallic lithium
target with protons of 1912~keV \cite{RaK88}. The experimental spectrum (solid line) 
represents a good approximation of a thermal spectrum for
$kT$~=~25 keV (dashed line). More detailed information on the angular dependence 
of the emitted neutron spectrum can be found in Ref. \cite{LiP75}.}
\label{fig_n_spec}
\end{figure}

The following discussion is therefore focused on the use of (p,n) reactions for 
neutron production. Among the different options for producing neutrons in the keV 
region, the $^7$Li(p,n)$^7$Be reaction with a threshold of 1.881 MeV is by far the most 
prolific. Near threshold, one can also take advantage of the fact that 
kinematically collimated neutrons can be produced in the energy range up to 100~keV.
A typical neutron spectrum derived on a regular basis during activation measurements at the 
Research Center Karlsruhe (FZK) is shown in Fig.~\ref{fig_n_spec}.

\subsection{Proton Accelerator}

Optimized, single ended Van-de-Graaff type accelerators are capable to produce 
pulsed proton beams with repetition rates between 100~kHz and 5~MHz. At present, 
the best performance of these machines allows one to achieve peak currents of 
14~mA at 1~MHz repetition rate and 0.7~ns pulse width, corresponding to average 
beam currents of up to 10~$\mu$A. Nevertheless, for easier comparison all 
values in the following tables and pictures are 
simply normalized to 10$^7$ protons, corresponding to a single
pulse at 1.6~$\mu$A beam intensity and 1~MHz repetition rate. Bombarding 
a thin metallic Li target just above the neutron production threshold, such a pulse
would provide a yield of 1.6$\times$10$^{-6}$~neutrons/proton, equivalent to 16~neutrons/pulse  
\cite{ReK02}. A proton current of 10~$\mu$A with 2~MeV energy implies a heat load at 
the target of only 20~J, which can be cooled with a jet of compressed air. If a more powerful
water cooling should be necessary, it needs to be designed to cool the target outside the
flight path of the neutrons towards the sample. This ensures an undisturbed neutron spectrum.

With with an optimized source, an integral source 
strength of 10${^8}$ neutrons/s is obtained in the energy range from 3 to 100~keV, 
which is the most relevant temperature regime for neutron capture nucleosynthesis. 
This neutron spectrum would be produced with a proton energy adjusted 30~keV above 
threshold and is kinematically collimated into an emission cone of 120~degrees 
opening angle. At higher proton energies this collimation effect vanishes and neutrons will 
be emitted into the entire solid angle. 

The potential of this neutron source can be illustrated by comparison with the 
LANSCE facility, the presently most powerful neutron source in the keV-region. 
For the new neutron capture experiment  DANCE the neutron flux between 3 and 
100~keV is expected to be 5$\times$10${^5}$ neutrons/s/cm$^2$, however at a more 
favorable duty cycle of 20~Hz.

Possible improvements of the current low energy accelerators appear feasible in 
several respects, for example by pulsing the extraction from the ion sources, by 
using pulsed dynamitrons, or 
by the addition of a proton storage ring, which would result in a further reduction 
of the repetition rate while keeping the average proton current.

\subsection{$\gamma$-Detector}

In neutron measurements on radioactive samples the $\gamma$-ray detection system 
has to meet a number of requirements:

1)	100\% efficiency for detecting small event rates and for background discrimination 
via the total $\gamma$-ray energy released per event,
2)	fast timing in order to achieve acceptable time-of-flight (TOF) resolution 
at short flight paths and to minimize pile-up effects,
3)	high granularity to reduce the count rate per module and to apply multiplicity cuts 
for background discrimination,
4)	good energy resolution to separate neutron captures on the sample from neutron captures on 
impurities with different total $\gamma$-ray energy, 
5)	low neutron sensitivity to avoid excessive backgrounds from scattered neutrons.

These requirements are best met by the 4$\pi$ BaF${_2}$ arrays at FZK \cite{WGK90a} and 
LANL \cite{WCD02} with 42 and 162 detector modules, respectively. BaF$_2$ turned out to 
be the most suited scintillator material because of the unique combination of 
fast timing, high efficiency, low neutron sensitivity, and good resolution in $\gamma$-ray energy.
An interesting alternative would be a 4$\pi$ array of CeF$_3$ scintillators \cite{HRK99}, but to
our knowledge, no large CeF$_3$ crystals are presently available.  

As outlined further on, the combination of a high granularity detector similar 
to the DANCE design and an advanced low energy neutron source appears to represent 
a promising concept for neutron capture experiments at RIA.

\subsection{Setup}

The proposed experimental setup is designed for an efficient TOF discrimination 
of the most important neutron related background components. The main components 
include

1)	a 4$\pi$ BaF${_2}$ array for detecting the prompt capture $\gamma$-rays, \\
2)	a high performance proton accelerator with the beam for neutron production 
ending in the center of the BaF${_2}$ array,\\
3)	a very short flight path of only 4~cm between neutron target and sample.\\

\begin{figure}
\begin{center}
\includegraphics[width=12cm]{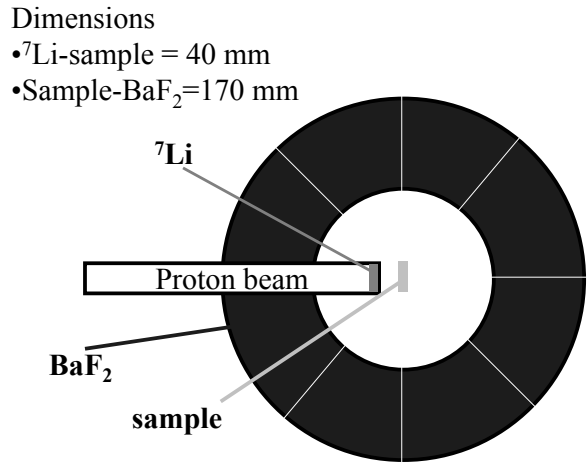}
\end{center}
\caption{Schematic sketch of the proposed setup. The segmentation of the 
crystals is indicated by white lines.}
\label{fig_set_schem}
\end{figure}

The schematic drawing of the proposed setup in Fig. \ref{fig_set_schem} indicates the underlying 
concept and shows the essential dimensions, i.e. an inner diameter of  the BaF${_2}$ 
array of 34~cm and the 4~cm distance between neutron target and sample.

\subsection{Expected TOF-spectra}

\begin{figure}
\begin{center}
\includegraphics[width=12cm]{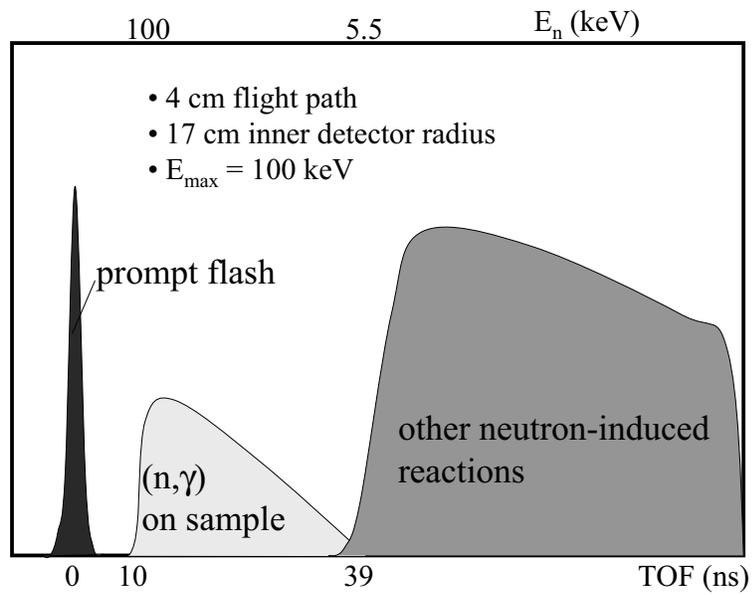}
\end{center}
\caption{Schematic TOF spectrum for the setup shown in Fig. \ref{fig_set_schem}.}
\label{fig_tof_schem}
\end{figure}

The schematic TOF spectrum in Fig. \ref{fig_tof_schem}  has been constructed under the following assumptions:\\
- 1~ns overall time resolution (proton pulse folded with BaF$_2$ resolution),\\
- 4~cm neutron flight path between $^7$Li target and sample,\\
- 17~cm inner radius of the BaF$_2$ array\\
- 100~keV maximum neutron energy.\\
It has been experimentally shown  that a time resolution of 1~ns can be achieved with the Karlruhe
4$\pi$~BaF$_2$ detector, which has a similar geometry like the one discussed here.

These parameters are sufficient to define the time and energy distributions of the 
setup sketched in Fig. \ref{fig_set_schem}. 
The neutron energy in keV at the sample position as a function of the neutron time of flight is given by: 

\begin{equation}
\label{tof_eq}
E_{neutron}[{\rm keV}]=522.6 \left(\frac{s[{\rm cm}]}{t[{\rm ns}]}\right) ^2~,
\end {equation}

where s is the flight path in cm and t the time of flight in ns.

\subsubsection{Neutron capture on the sample}
 
The useful TOF window between the prompt $\gamma$-flash and the onset of the background from 
scattered neutrons can be calculated by the following steps:

1)	The 4~cm flight path between neutron target and sample is determined by requiring a 
clear separation of 10 ns in TOF between the prompt $\gamma$-flash due to the impact of the 
proton beam and the arrival time of the fastest neutrons (100~keV) at the sample.\\
2)	The background from scattered neutrons starts when the fastest neutrons arrive at 
the scintillator. These neutrons originate either from scattering at the Li target or 
at the sample. This occurs at a TOF of 39~ns when primary neutrons of 5.5~keV energy 
interact with the sample.

Accordingly, the neutron energy range from 5.5~keV to 100~keV is essentially free of 
time-dependent background, corresponding exactly to the most relevant part for 
cross section measurements of astrophysical interest. The 4~cm flight path implies 
a reduction in neutron flux compared to the total source strength by about a factor 
of 10, significantly less compared to the solid angle, because the neutron 
distribution is strongly forward peaked due to reaction kinematics. This reduction 
may be compensated by the possible improvements in proton beam intensity as mentioned 
above. Though the TOF measurement is hampered by the short flight path, the above 
parameters yield a neutron energy resolution of 12\% at 30 keV, assuming that the energy
resolution is determined by the TOF resolution. This holds true, if the uncertainty of the flight path
is less than 6\%. 

At energies above 120~keV, neutron emission is no longer collimated by reaction 
kinematics, but becomes increasingly isotropic. In addition, the TOF scale sketched 
in Fig. \ref{fig_tof_schem} is more and more compressed due to the higher neutron velocities, 
so that separation of the three components becomes more difficult. Measurements 
in this energy range have to rely on the production of quasi mono-energetic neutrons, 
where the energy is defined by the thickness of the $^7$Li target. For neutrons of 
500~keV the flight path must be increased to 9~cm in order to maintain the 10~ns 
separation between the prompt $\gamma$-flash and neutron capture events. Assuming, the 
inner radius of the scintillator would stay at 17~cm, the background from neutron interactions 
with the scintillator starts 
already 17~ns past the $\gamma$-flash. Though the useful TOF window is reduced to 7~ns in 
this case, it still corresponds to an energy interval between 500 and 200~keV. In 
other words, the method works in principle also at higher energies since the thickness 
of the Li targets can be defined within  a few keV.

Energies below 5.5~keV could be obtained by a combination of shorter flight path, increased inner
diameter of the ball, and reduced maximum neutron energy. The latter is possible by reducing the incident proton energy
or by bombarding a different neutron production target, like $^{18}$O. 

\subsubsection{The prompt $\gamma$-flash}

\begin{figure}
\begin{center}
\includegraphics[width=12cm]{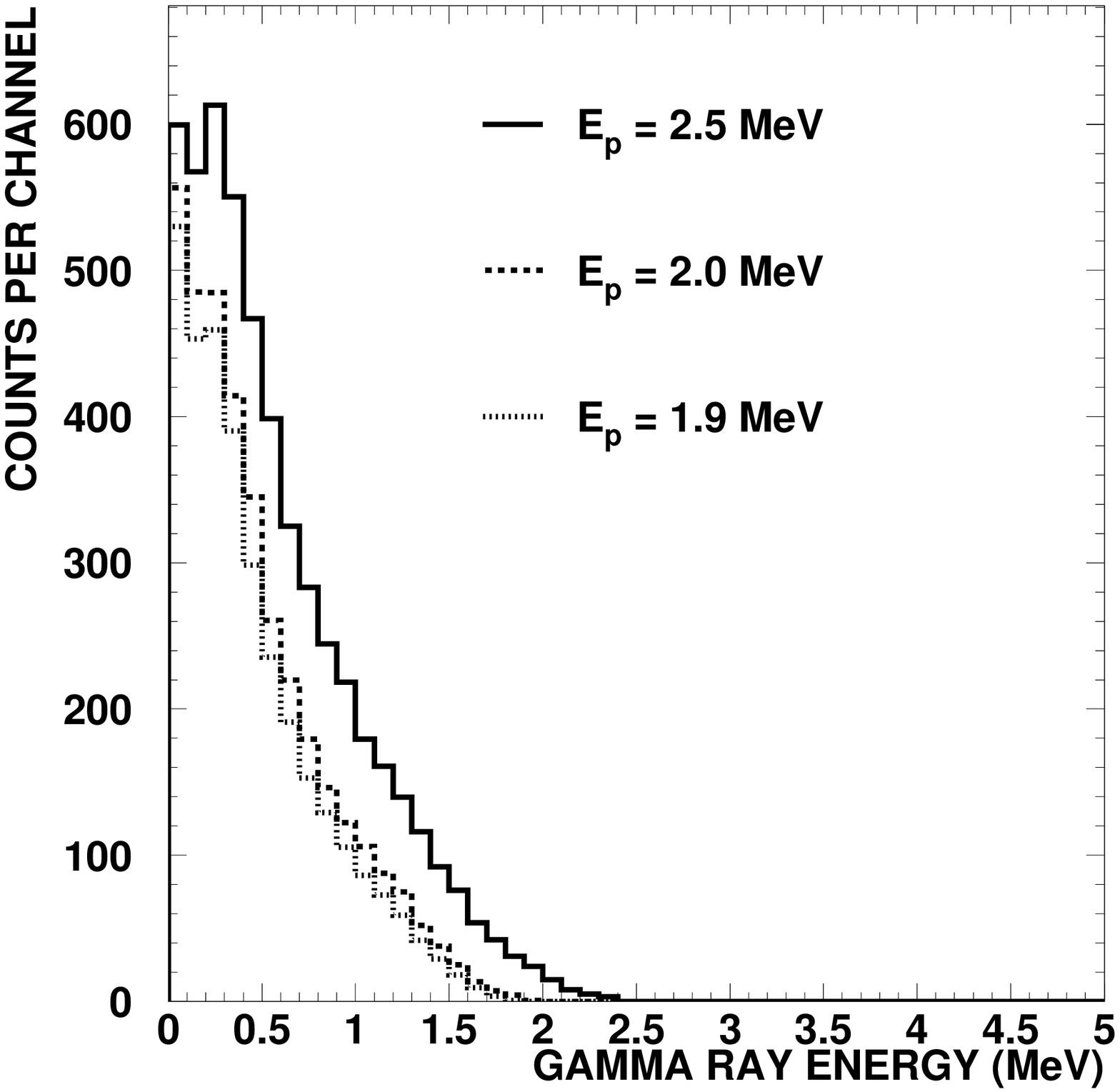}
\end{center}
\caption{Interaction of 10$^7$ protons with a 0.2 mm Bi target. The histogram binning is 100 keV/channel.}
\label{fig_bi}
\end{figure}

\begin{figure}
\begin{center}
\includegraphics[width=12cm]{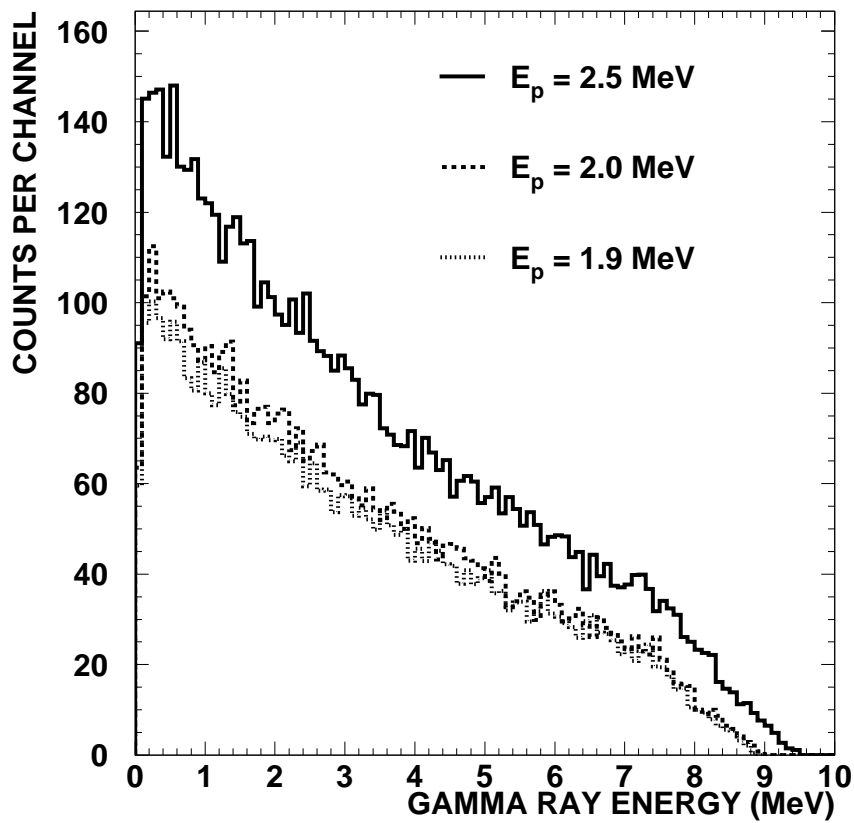}
\end{center}
\caption{Interaction of 10$^7$ protons with a 0.5 mm Cu target. Notice the 
 different x-axis compared to Figs. \ref{fig_bi} and \ref{fig_li}. The histogram binning is 100 keV/channel.}
\label{fig_cu05}
\end{figure}

\begin{figure}
\begin{center}
\includegraphics[width=12cm]{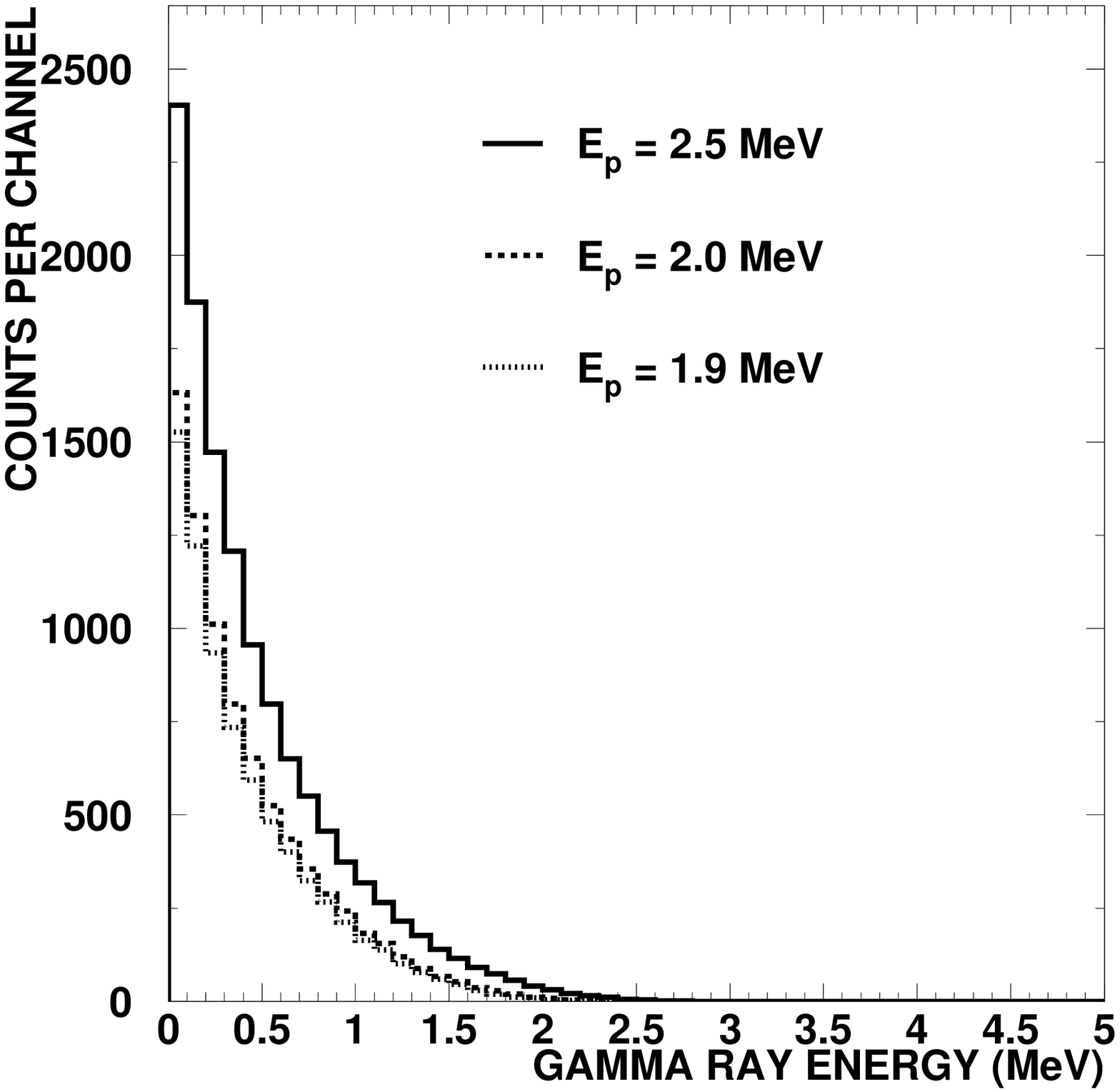}
\end{center}
\caption{Interaction of 10$^7$ protons with a 0.5 mm Li target. The histogram binning is 100 keV/channel.}
\label{fig_li}
\end{figure}

\begin{figure}
\begin{center}
\includegraphics[width=12cm]{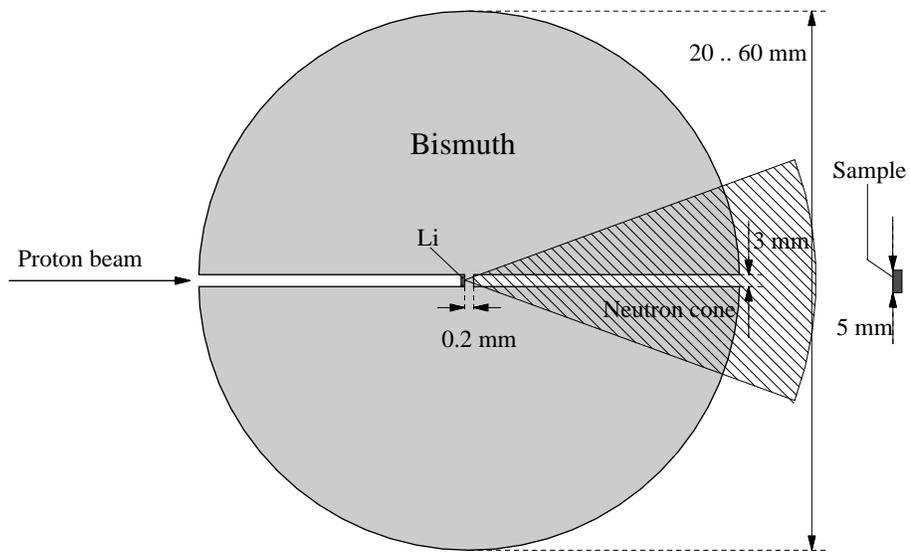}
\end{center}
\caption{Schematic view of the simulated spherical Bi targets. The drawing is not on scale. Neutrons 
are produced via $^7$Li(p,n) at a thin lithium layer inside the bismuth absorber sphere.}
\label{fig_bi_sphere}
\end{figure}

\begin{figure}
\begin{center}
\includegraphics[width=12cm]{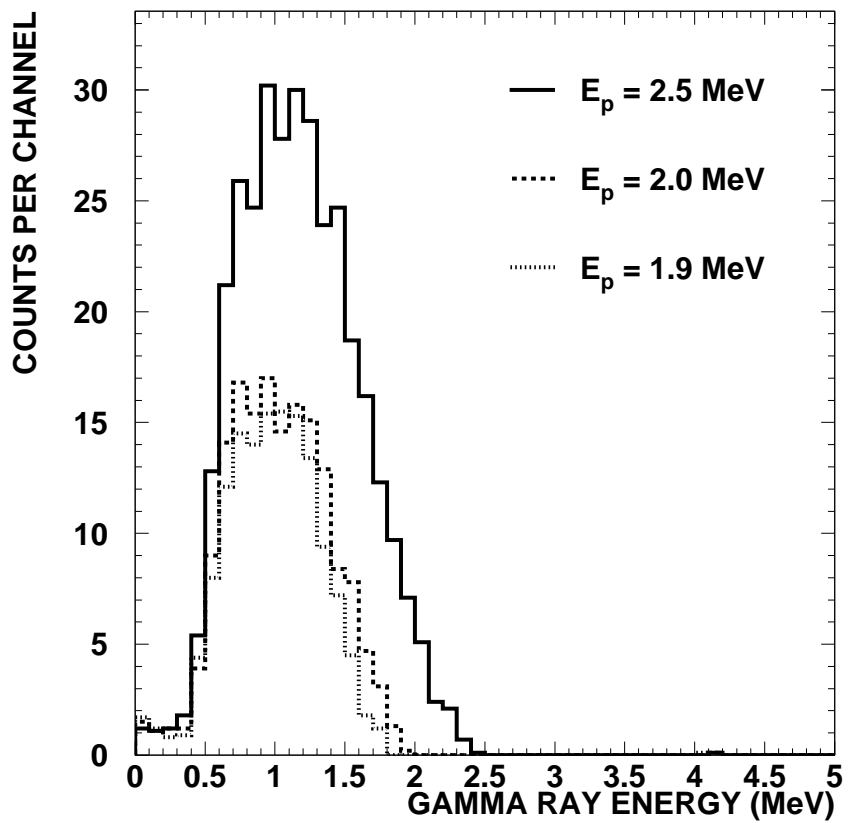}
\end{center}
\caption{Interaction of 10$^7$ protons with an 0.2 mm Bi target with a surrounding Bi sphere of 3 cm radius
(see Fig. \ref{fig_bi_sphere}). The histogram binning is 100 keV/channel.}
\label{fig_bi_3cm}
\end{figure}

The origin of the so-called prompt $\gamma$-flash are interactions of the proton pulse 
and of the produced neutrons with the lithium target and the backing. Most of the energy
of the protons will be deposited in the backing material. Photons will be produced due to bremsstrahlung,
inelastic scattering, and nuclear reactions. In order to reduce the number of photons and the total
energy released during a single pulse, a set of simulations have been carried out for different 
backing materials (Table \ref{tab_back_1}). The corresponding spectra are shown in Figs. 
\ref{fig_bi} to \ref{fig_li}. As a general trend one finds that the total number of emitted $\gamma$-rays as well as the
averaged energy per $\gamma$-ray decreases with increasing atomic number of the backing material, if no nuclear 
interactions are involved (which is true for all investigated backing materials except copper).
But even for a backing made of bismuth, the heaviest stable element, the total energy released during one pulse
of 10$^7$ protons is more than 1~GeV. Without any further precautions, the detector
would be blinded for at least 10~ns.\\
One way of reducing this $\gamma$-flash is passive shielding. 
A possible backing and absorber combination made of bismuth is sketched in Figure~\ref{fig_bi_sphere}. 
Bismuth is a good choice for several reasons. First, Table \ref{tab_back_2} shows that even though the density 
of bismuth is significantly smaller than that of tungsten, the 
energy deposited in the surrounding detector is sightly less. Second, 
bismuth has only one stable isotope with a small average (n,$\gamma$) cross section
of a few mb at keV energies \cite{BBK00}, and a small neutron capture Q-value of only 4 MeV, 
in contrast to tungsten and most other elements. 
Furthermore, there is no inelastic scattering to be expected for the discussed 
neutron energy range, since the first excited state 
of $^{209}$Bi is at 896.3~keV \cite{Fir96}. \\
This means that most of the neutrons, which will be emitted 
into such an absorber sphere, will be scattered and eventually leave the sphere without nuclear interaction. Only a very small 
part will be captured and release at most 4~MeV, which is well below the Q-value of most of the isotopes of interest. Depending 
on the thickness of the absorber, the energy deposition due to the $\gamma$-flash will be reduced by a factor of 2 to 10 and 
the averaged energy per photon is increased (the spectrum is becoming harder).
The resulting spectra for a Bi absorber with 3 cm radius are shown in Fig. \ref{fig_bi_3cm}. \\
If a bismuth absorber would not be possible due to technical reasons,
tungsten would be the second choice. Therefore, Table \ref{tab_back_2} 
contains the results of the simulations also 
for a tungsten absorber.

\subsubsection{Other neutron-induced reactions}

\begin{figure}
\begin{center}
\includegraphics[width=12cm]{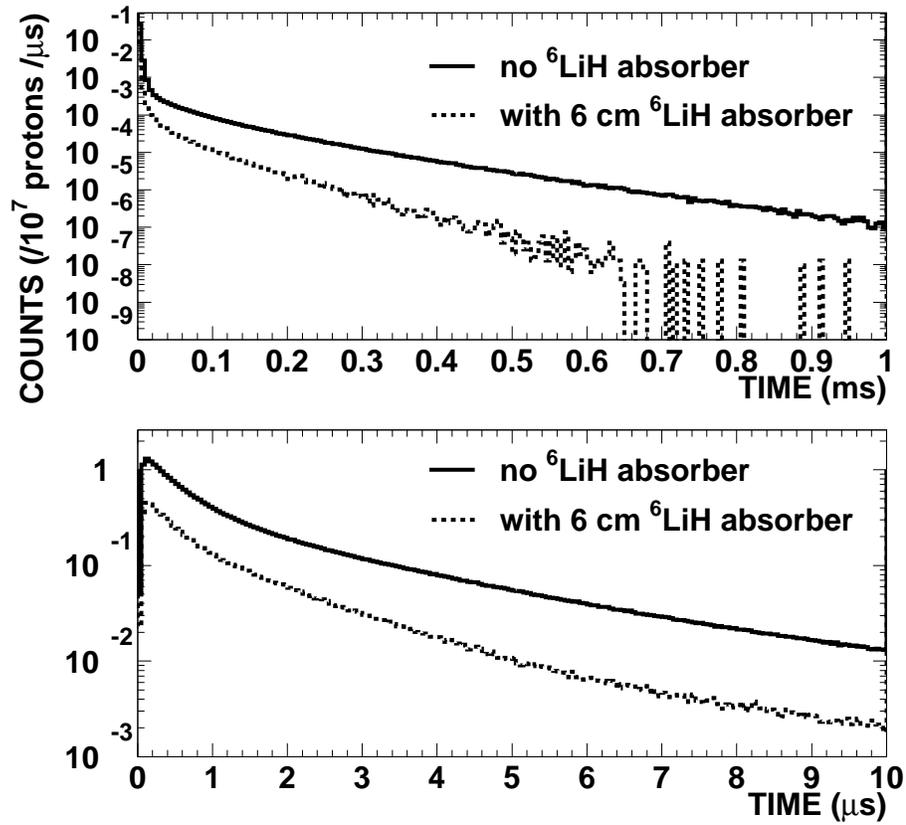}
\end{center}
\caption{TOF spectrum for events due to neutron capture in the surrounding material (crystal ball etc.) 
The spectra have been simulated including a $^6$LiH neutron absorber shell of 6~cm thickness around the Li target. The
total number of neutrons corresponds to 10$^7$ protons hitting the metallic Li target.}
\label{nn_tof_ms}
\end{figure}

\begin{figure}
\begin{center}
\includegraphics[width=12cm]{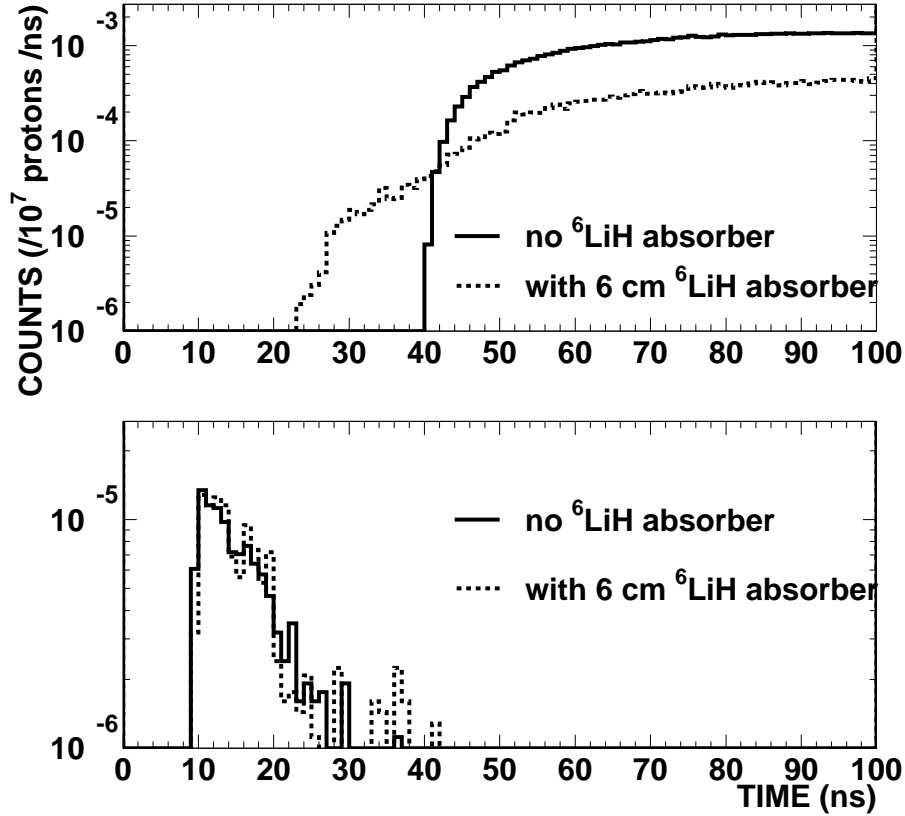}
\end{center}
\caption{TOF spectra for events due to neutron capture in the surrounding material (top) and in an 
0.2 mm thick gold sample (bottom).
The spectra have been simulated with and without a $^6$LiH neutron absorber shell of 6~cm thickness around the Li target. 
Accordingly, $^1$H(n,$\gamma$)$^2$H reactions in the $^6$LiH absorber cause the background of scattered neutrons to start at earlier 
TOF. This component can be distinguished because its Q-value of 2.2~MeV is considerably 
lower than the one for gold capture events. As expected, the $^6$LiH absorber does not affect the gold spectrum. 
The total number of neutrons corresponds to 10$^7$ protons hitting the metallic Li target.}
\label{nn_tof_ns}
\end{figure}

\begin{figure}
\begin{center}
\includegraphics[width=12cm]{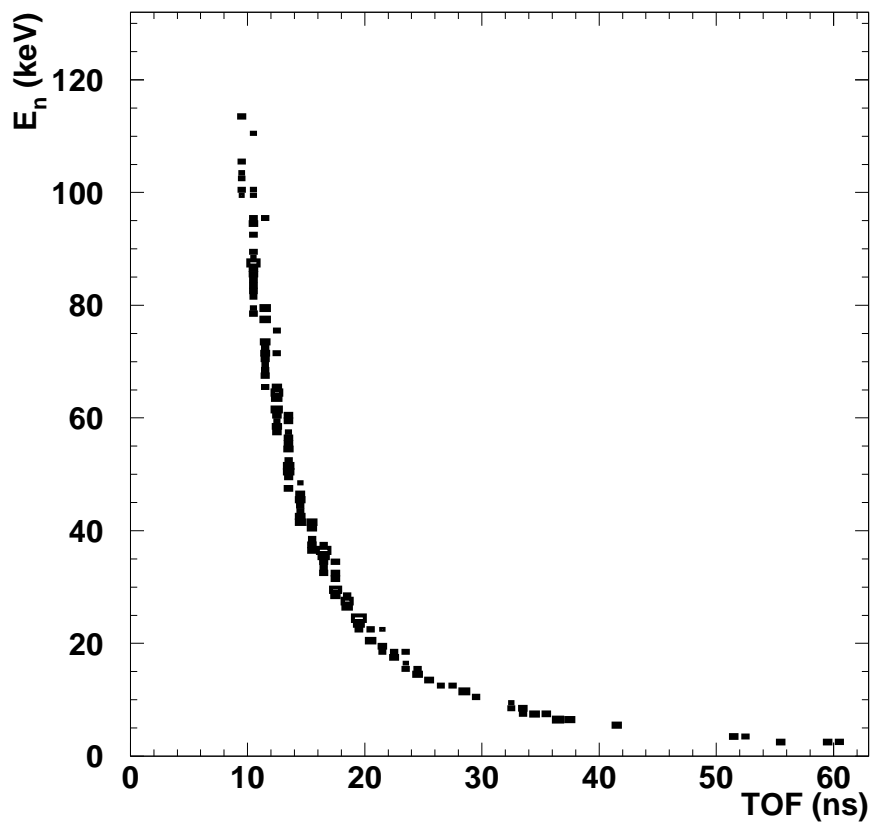}
\end{center}
\caption{Neutron energy versus TOF for capture events on the gold sample. 
All events are following the expected quadratic time-to-energy
dependence determined by Eq. (\ref{tof_eq}), thus confirming that the 
capture contribution of scattered neutrons in the sample is negligible 
(see text).}
\label{en_tof}
\end{figure}

Most of the neutron-induced reactions can be discriminated via TOF, since the neutrons have
to travel at least the distance from the center of the array to the scintillator. But there are three
additional sources of background, which could contribute at times, when neutron 
captures at the sample are registered. \\
First, neutrons from the preceding pulse may be captured in the detector with 
sufficient time delay to cause background events at times,
when only captures on the sample are to be expected. Since a repetition
rate of 1~MHz corresponds to a pulse spacing of 1~$\mu$s, this so-called wrap-around effect is not negligible as illustrated in 
Figs. \ref{nn_tof_ms} and \ref{nn_tof_ns}. If the experiment has to be carried out at a repetition rate of 1~MHz, 
further reduction of the background from wrap-around neutrons is required.
Therefore, the same simulation has been repeated including 
a spherical $^6$LiH shell with an inner radius of 10.5~cm and an outer radius of 16.5~cm. This shell acts as an efficient 
neutron absorber via $^6$Li(n,$\alpha$), while it is very transparent to $\gamma$-rays (see caption of Fig. \ref{nn_tof_ns} for details). \\
Depending on the neutron TOF, the wrap-around background is reduced by up to two orders of magnitude. 
If the neutron background level after 1~$\mu$s of waiting
time would still be unacceptably high, the number of wrap-around neutrons could
be significantly reduced by choosing a lower repetition rate. \\  
Second, neutrons can get scattered at the different absorber materials and may reach the gold target with a wrong 
energy-TOF correlation. Fig. \ref{en_tof} shows that this effect is negligible. \\
Third, neutron captures on the shielding for the prompt $\gamma$-flash, on the neutron absorber, and on the support of 
the lithium target have to be considered as well. Depending on the
materials used, such events can be discriminated via the total energy released. 
For example, the Q-value of the $^{209}$Bi(n,$\gamma$)$^{210}$Bi reaction is
4.606~MeV, about 2~MeV below the neutron separation energies of most isotopes near stability. 
This background reduction based on cuts on the Q-value is discussed in more detail
in the following section.

\subsection{Additional Background Discrimination - Expected Energy Spectra}

\begin{figure}
\begin{center}
\includegraphics[width=12cm]{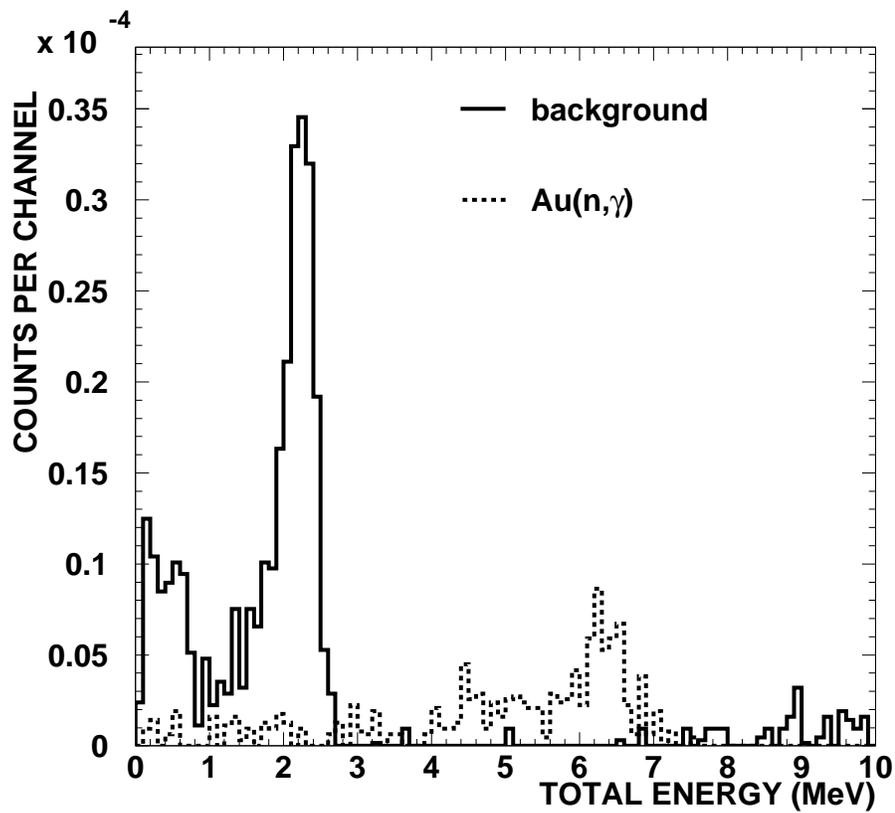}
\end{center}
\caption{Total energy deposited in the 4$\pi$ array due to capture in gold (dashed 
line) and background reactions (solid curve) for the TOF spectrum shown in 
Fig. (\ref{nn_tof_ns}). The background contains captures on $^1$H and $^6$Li as 
well as inelastic scattering. An energy cut between 3 and 8 MeV will eliminate 
most of this background. The energy spectrum is shown for a TOF cut of 0-40~ns, 
which means that captures on the scintillator material are excluded.
The total number of neutrons corresponds to 10$^7$ protons hitting the metallic 
Li target. The histogram binning is 100 keV/channel.}
\label{nn_tof_ns_40}
\end{figure}

While almost all beam-related background components can be discriminated via TOF, 
backgrounds due to neutron interactions with the $^6$LiH absorber, the $\gamma$-ray shield,
the sample backing, inelastic neutron scattering at the sample, sample impurities and from the 
radioactivity of the sample or due to activations in the detector have to be considered as well. 
Fig. \ref{nn_tof_ns_40}
illustrates the situation for the standard setup of an 0.2~mm thick gold sample with 0.5 cm radius, and a 6~cm thick
$^6$LiH shell around the sample. Events due to neutron capture on gold can 
clearly be identified via their total deposited energy. These spectra were derived by using  only
the fast component of the BaF$_2$ scintillation light and a correspondingly lower 
resolution of the 4$\pi$ array. \\
In measurements on very small samples it is not always possible to keep neutron captures 
on the backing or on sample impurities at a negligible level. Since these neutron 
captures are taking place at the sample position, they exhibit the same time structure as 
captures on the isotope of interest. However, the almost 100\% detection efficiency 
for $\gamma$-rays in combination with the high granularity of the proposed detector provides a 
variety of possibilities for discriminating these background events. First, the total energy 
(Q-value) of the reaction depends on the respective isotope. Even though only the fast 
component of the BaF$_2$ crystals, containing about 20\% of the emitted light, can be used 
in the proposed setup, the energy resolution at a Q-value of 7~MeV, which is characteristic 
for the (n,$\gamma$)-reactions of interest, is about 400~keV. If the background is dominated by a 
single isotope, it can be discriminated by the total $\gamma$-ray energy. Second, the average 
multiplicity of the prompt capture $\gamma$-ray cascade of the isotope under study is usually different 
from background events. This information is recorded thanks to the high granularity of 
the BaF$_2$ array \cite{HRF01, RHK01}. \\
Both options are particularly efficient with respect to the background due to the radioactivity 
of the sample or activated detector parts as well as inelastic scattering on the sample. 
In most cases, the total $\gamma$-ray energy released in radioactive decay is less than 2~MeV, 
significantly below average neutron separation energies and is often restricted to 
multiplicities of less than~3. Inelastic neutron scattering produces a $\gamma$-ray, potentially  
at the same time as neutron captures
would occur. Since the emitted $\gamma$-ray energy needs to be smaller than the neutron energy, 
it will be below 1~MeV and the multiplicity will be one for most cases. Therefore, both background 
components can be easily discriminated from neutron captures with typical Q-values of a few MeV.\\
Any remaining components, e.g. due to ambient background, can be determined
by means of background runs without sample or with an empty sample backing. 
 
\section{Comparison with Existing Facilities}

At present, the most sensitive setup for (n,$\gamma$) measurements in the keV 
region is the DANCE project at 
LANSCE/LANL \cite{WCD02}, which is designed for samples of about 1~mg in mass. 
Therefore, the proposed setup will only be compared with this facility. 
 
The total number of neutrons per second is directly correlated to the amount of 
sample material necessary to obtain the required statistics in a given time. 
Depending on the actual solution, the proposed setup is able to deliver about $10^7$ neutrons/s
to the sample position, which is 20~times more neutrons than available at DANCE. \\
Another important aspect concerns the signal-to-background ratio. As mentioned above, 
the radioactivity of the sample is one background component, that can not be discriminated 
via time of flight. The number of $\gamma$-rays per time produced via radioactive decay 
compared to the number of neutron captures per time is, therefore, important. A number of different
neutron sources has been investigated by P. Koehler, using essentially this figure-of-merit 
\cite{Koe01b}.
The following equation shows the ratio of neutron captures per time at the possible RIA facility 
and DANCE, where $\Phi$ stands for the neutron flux, TOF for the lengths of the time-of-flight 
interval of interest, which is proportional to the flight path, and f for the repetition rate. 

\begin{eqnarray*}
\frac{RIA}{DANCE} & = & \frac{\Phi_{RIA}}{\Phi_{DANCE}} \cdot \frac{TOF_{DANCE}}{TOF_{RIA}} 
                        \cdot \frac{f_{DANCE}}{f_{RIA}} \\
                  & = & \frac{10^7}{5 \cdot 10^5} \cdot \frac{20}{0.04} \cdot \frac{20}{10^6} \\
				  & = & 0.2
\end{eqnarray*}

This implies, that the instantaneous capture rates at both facilities are not too 
different and can become comparable with the possible improvements of the proton
beam intensity. 

\section{Count rate and half-life estimates}

Assuming a desired number of 10$^3$ counts, a neutron flux of 10$^7$ neutrons/s/cm$^{2}$, an averaged 
neutron capture cross section of 200~mb, and a measuring time 
of 2$\times$10$^6$~s, a sample of 2.5$\times$10$^{14}$ atoms would be sufficient for the proposed setup. \\
Close to the valley  of stability, RIA will reach beam intensities of 10$^{11}$ to 
10$^{12}$ particles/s. Accordingly, such samples 
could be produced within a few hours, presumably even in parallel to normal operation 
if the collection can run in parasitic mode. \\
Because of the excellent time resolution of the proposed setup, $\gamma$-rays 
in the critical  
energy range above 100~keV can be handled up to a total rate of 20~MHz. 
The limit of 20 MHz for the total rate seen by the array corresponds to
a conservatively estimated count rate of $<$1~MHz per detector, even 
if cross talk and cascade transitions are taken into account.\\
The minimum accessible half-life can be estimated by:

\begin{equation}
t_{1/2}=\frac{\rm ln 2} {\lambda}={\rm ln 2} \cdot \frac{N}{A}={\rm ln 2} \cdot \eta \frac{N}{A_\gamma}\approx
\eta \cdot 10^7 {\rm s}~,
\end{equation}

where t$_{1/2}$	means the half-life of the isotope under investigation, $\lambda$ the decay constant,
N the number of sample atoms (2.5$\times$10$^{14}$), A the sample activity (decays per second), 
$\eta$ the relative intensity per decay for $\gamma$-rays above 100 keV, and 
A$_\gamma$=~$\eta\cdot$A~=~2$\times$10$^7$~s$^{-1}$ the acceptable $\gamma$ activity 
above 100 keV.\\

Accordingly, for cases with $\eta~\le~0.1$, isotopes with half-lives down to 10~d can be investigated, consistent 
with the number of sample atoms available from RIA.

\section{Conclusions}

This paper presents a possible way for neutron capture measurements on short-lived isotopes
using an existing total absorption
calorimeter close to a RIA type facility. While an optimized design will depend on the specific 
features of the actual facility, this paper attempts to demonstrate that neutron 
capture measurements in the keV~range with minimal sample masses of $\approx$10$^{15}$ 
atoms are feasible in principle. Depending on the number of $\gamma$-rays 
emitted per decay, this implies a minimum half-life of the isotope under investigation 
of only 10 d. Prominent examples, which became accessible with this facility, would 
be $^{60}$Co, $^{110}$Ag, $^{137}$Cs, $^{147}$Nd, $^{146}$Pm, $^{148}$Pm, $^{170}$Tm for 
analyses of s-process branchings as 
well as for a variety of isotopes of relevance for explosive nucleosynthesis in the 
r and p processes. Correspondingly, measurements on technologically relevant cross sections could 
be performed as well, e.g. on $^{237}$U. 

\ack{
This work has benefited from the use of the Los Alamos Neutron Science Center at the Los Alamos 
National Laboratory. This facility is funded by the US Department of Energy and operated by the 
University of California under Contract W-7405-ENG-36.
}


\newcommand{\noopsort}[1]{} \newcommand{\printfirst}[2]{#1}
  \newcommand{\singleletter}[1]{#1} \newcommand{\swithchargs}[2]{#2#1}

\begin{table}
 \caption{Number of $\gamma$-rays produced by the interaction of 10$^7$ protons with different backings
 (corresponing to one pulse at $\approx1.6~\mu$A and 1~MHz)}
   \label{tab_back_1}
   \begin{tabular}{cccccc}
    Backing & Thickness & Proton energy & Total energy & Number of     & Average $\gamma$-energy \\
    Backing & (mm)      & (MeV)         & (GeV)        & $\gamma$-rays & (MeV) \\
    \hline
    Bi    & 0.2 & 1.9 & 1.60 & 3270 & 0.496\\
          &     & 2.0 & 1.88 & 3620 & 0.518\\
          &     & 2.5 & 3.28 & 5210 & 0.629\\	
    \hline		       
    Pt    & 0.2 & 1.9 & 2.04 & 3540 & 0.577\\
          &     & 2.0 & 2.27 & 3790 & 0.598\\
          &     & 2.5 & 3.83 & 5410 & 0.708\\
    \hline
    W     & 0.2 & 1.9 & 2.50 & 3680 & 0.679\\
          &     & 2.0 & 2.75 & 3930 & 0.698\\
          &     & 2.5 & 4.42 & 5500 & 0.803\\	
    \hline	   
    Ta    & 0.2 & 1.9 & 1.75 & 3350 & 0.522\\
          &     & 2.0 & 2.05 & 3660 & 0.560\\
          &     & 2.5 & 3.70 & 5270 & 0.702\\	
    \hline		      
    Cu    & 0.5 & 1.9 & 12.5 & 4050 & 3.08\\
          &     & 2.0 & 13.4 & 4370 & 3.08\\
          &     & 2.5 & 20.0 & 6210 & 3.21\\	
    \hline		       
    Cu    & 2.0 & 1.9 & 12.2 & 3950 & 3.08\\
          &     & 2.0 & 13.1 & 4230 & 3.09\\
          &     & 2.5 & 19.6 & 6080 & 3.21\\	
    \hline		       
    Li    & 0.5 & 1.9 & 3.45 & 7360 & 0.468\\
          &     & 2.0 & 3.83 & 8020 & 0.478\\
          &     & 2.5 & 6.45 & 12300 & 0.524\\	
    \hline		       
    Be    & 0.5 & 1.9 & 3.12 & 6420 & 0.485\\
          &     & 2.0 & 3.46 & 6950 & 0.498\\
          &     & 2.5 & 5.66 & 10100 & 0.561\\

   \end{tabular}
\end{table}

\begin{table}
 \caption{Number of $\gamma$-rays produced by the interaction of 10$^7$ protons with different W and Bi backings surrounded by
 a spherical absorber (see Fig. \ref{fig_bi_sphere}).}
   \label{tab_back_2}
   \begin{tabular}{ccccccc}
    Backing & Thickness & Sphere Radius   &Proton energy & Total energy & Number of     & Average $\gamma$-energy \\
    Backing & (mm)          & (mm)       & (MeV)        & (MeV)        & $\gamma$-rays & (MeV) \\
    \hline
    W     & 0.2 & 30  & 1.9 & 155 & 93.5 & 1.66\\
          &     &     & 2.0 & 191 & 110 & 1.73\\
          &     &     & 2.5 & 309 & 183 & 1.69\\	
    \hline
    W     & 0.2 & 20  & 1.9 & 325 & 255 & 1.27\\
          &     &     & 2.0 & 354 & 265 & 1.33\\
          &     &     & 2.5 & 631 & 459 & 1.38\\				
    \hline
    W     & 0.2 & 10  & 1.9 & 805 & 783 & 1.03\\
          &     &     & 2.0 & 934 & 898 & 1.05\\
          &     &     & 2.5 & 1610 & 1400 & 1.15\\			    
    \hline
    Bi    & 0.2 & 30  & 1.9 & 140 & 141 & 0.988\\
          &     &     & 2.0 & 170 & 165 & 1.03\\
          &     &     & 2.5 & 394 & 335 & 1.18\\			  		  
    \hline
    Bi    & 0.2 & 20  & 1.9 & 274 & 297 & 0.924\\
          &     &     & 2.0 & 316 & 334 & 0.950\\
          &     &     & 2.5 & 735 & 683 & 1.08\\			  		  
    \hline
    Bi    & 0.2 & 10  & 1.9 & 591 & 753 & 0.785\\
          &     &     & 2.0 & 699 & 857 & 0.815\\
          &     &     & 2.5 & 1460 & 1560 & 0.931\\			  		  
   \end{tabular}
\end{table}

\clearpage

\vspace*{5mm}

\end{document}